\documentclass[10pt,letterpaper]{article}
\usepackage[top=0.85in,left=2.75in,footskip=0.75in]{geometry}

\usepackage{changepage}

\usepackage[utf8x]{inputenc}

\usepackage{textcomp,marvosym}

\usepackage{amsmath,amssymb}

\usepackage{cite}

\usepackage{nameref,hyperref}

\usepackage[right]{lineno}

\usepackage{microtype}
\DisableLigatures[f]{encoding = *, family = * }

\raggedright
\usepackage[aboveskip=1pt,labelfont=bf,labelsep=period,justification=raggedright,singlelinecheck=off]{caption}

\makeatletter
\renewcommand{\@biblabel}[1]{\quad#1.}
\makeatother

\date{}

\usepackage{lastpage,fancyhdr,graphicx}
\usepackage{epstopdf}
\fancyheadoffset[L]{2.25in}
\fancyfootoffset[L]{2.25in}
\begin{document}
\vspace*{0.2in}

\begin{flushleft}
{\Large
\textbf\newline{A General Approach for Predicting the Behavior of the Supreme Court of the United States} 
}
\newline
\\
Daniel Martin Katz$^{1,2*}$, 
Michael J Bommarito II$^{1,2}$, 
Josh Blackman$^{3}$
\\
\bigskip
\textbf{1} Illinois Tech - Chicago-Kent College of Law, Chicago, IL, USA
\\
\textbf{2} CodeX - The Stanford Center for Legal Informatics, Stanford, CA, USA
\\
\textbf{3} South Texas College of Law Houston, Houston, TX, USA
\\
\smallskip
\smallskip
\smallskip
* E-mail: dkatz3@kentlaw.iit.edu 
\bigskip

%
%






\end{flushleft}
\section*{Abstract}
Building on developments in machine learning and prior work in the science of judicial prediction, we construct a model designed to predict the behavior of the Supreme Court of the United States in a generalized, out-of-sample context.  To do so, we develop a time evolving random forest classifier which leverages some unique feature engineering to predict more than 240,000 justice votes and 28,000 cases outcomes over nearly two centuries (1816-2015). Using only data available prior to decision, our model outperforms null (baseline) models at both the justice and case level under both parametric and non-parametric tests.  Over nearly two centuries, we achieve 70.2\% accuracy at the case outcome level and 71.9\% at the justice vote level. More recently, over the past century, we outperform an \textit{in-sample optimized} null model by nearly 5 \%.  Our performance is consistent with, and improves on the general level of prediction demonstrated by prior work;  however, our model is distinctive because it can be applied out-of-sample to the entire past and future of the Court, not a single term.  Our results represent an important advance for the science of quantitative legal prediction and portend a range of other potential applications. 



\section*{Introduction}
As the leaves begin to fall each October, the first Monday marks the beginning of another term for the Supreme Court of the United States.  Each term brings with it a series of challenging, important cases that cover legal questions as diverse as tax law, freedom of speech, patent law, administrative law, equal protection, and environmental law.  In many instances, the Court’s decisions are meaningful not just for the litigants \textit{per se}, but for society as a whole.

Unsurprisingly, predicting the behavior of the Court is one of the great pastimes for legal and political observers.  Every year, newspapers, television and radio pundits, academic journals, law reviews, magazines, blogs, and tweets predict how the Court will rule in a particular case.  Will the Justices vote based on the political preferences of the President who appointed them or form a coalition along other dimensions?  Will the Court counter expectations with an unexpected ruling?

Despite the multitude of pundits and vast human effort devoted to the task, the quality of the resulting predictions and the underlying models supporting most forecasts is unclear.  Not only are these models not backtested historically, but many are difficult to formalize or reproduce at all.  When models are formalized, they are typically assessed ex post to infer causes, rather than used ex ante to predict future cases.  As noted in \cite{bib1}, ``the best test of an explanatory theory is its ability to predict future events.  To the extent that scholars in both disciplines (social science and law) seek to explain court behavior, they ought to test their theories not only against cases already decided, but against future outcomes as well.''

Luckily, the Court provides a new opportunity to test each year.  Thousands of petitioners annually appeal their cases to the Supreme Court.  In most situations, the Court decides to hear a case by granting a petition for a writ of certiorari. If that petition is granted, the parties then submit written materials supporting their position and later provide oral argument before the Court. After considering the case, each participating Justice ultimately casts his or her vote on whether to affirm or reverse the status quo (typically seen through the lens of a decision by the lower court or special master). Over the last decade, the Court has issued between 70-90 opinions per term for an average of approximately 700 Justice votes per term.  

While many questions could in principle, be evaluated, the Court's decisions offer at least two discrete prediction questions: 1) will the Court as a whole affirm or reverse the status quo judgment and 2) will each individual Justice vote to affirm or reverse the status quo judgment? 

In this paper, we describe a prediction model answering these two questions as guided by three modeling goals: generality, consistency, and out-of-sample applicability. Building on developments in machine learning and the prior work of \cite{bib1},\cite{bib2} and\cite{bib3}, we construct a model to predict the voting behavior of the Court and its Justices in a generalized, out-of-sample context.  As inputs, we rely upon the Supreme Court Database (SCDB) and some derived features generated through feature engineering.  Our model is based on the random forest method developed in \cite{bib4}.  We predict nearly two centuries of historical decisions (1816-2015) and compare our results against multiple null (baseline) models.  

Using only data available prior to decision, our model outperforms all baseline models at both the Justice and Court observation level under both parametric and non-parametric tests.  This performance is consistent with, and improves on the general level of prediction demonstrated by prior work;  however, our model is distinctive because it can be applied out-of-sample to the entire past and future of the Court, not just a single term.  Finally, our conclusion suggests areas for future improvement and collaboration.  Our results represent a significant advance for the science of quantitative legal prediction and portend a range of potential applications, such as those described in\cite{bib5}.  

\section*{Research Principles and Prior Work}

In this section, we describe the principles guiding our model construction and how we conducted our testing in light of prior work on the topic.

\subsection*{Generality}
Leveraging the early work of \cite{bib6}, both \cite{bib1} and \cite{bib3} developed a classification tree model which was designed to predict the behavior of Supreme Court Justices for the 2002-2003 Supreme Court term. Their work represents a seminal contribution to the science of legal forecasting as their classification tree models not only performed well in absolute terms, but also matched or outperformed a number of subject matter experts.  

Despite its contribution to the field, however, the approach undertaken in \cite{bib1} and \cite{bib3} was limited in several important ways. For example, their model construction is only applicable to a single “natural court” with full participation, i.e., cases where all of a specific set of Justices are sitting.  The natural court tested in their paper, following Justice Stephen G. Breyer’s appointment in 1994, was one of the longest periods without personnel changes on the Court, providing their models with an unusually large training sample.  It is not possible, however, to evaluate their model in periods prior to 1994 or after 2005 following the replacements of Chief Justice William H. Rehnquist and Justices Sandra Day O’Connor, David H. Souter, and John Paul Stevens.  As a result of these issues, the performance and nature of the model cannot necessarily be generalized to all Supreme Court cases during their test period, let alone cases before or after their tested natural court.

Our first principle, generality, is based on these observations.  As the composition of the Court changes case-by-case or term-by-term, either through recusal, retirement, or death, a prediction model should continue to generate predictions.  The properties and performance of a prediction model should also be able to be studied across time and “abnormal” circumstances (e.g., cases with original jurisdiction or fewer than nine Justices). Therefore, our goal is to construct a model that is general - that is, a model that can learn \textit{online}, in a manner similar to online learning models described in \cite{bib7} and \cite{bib8}.
  
\subsection*{Consistency}
Second, we prefer the model to have consistent performance across time, case issues, and Justices.  Similar to our motivation for generality, existing models have had significantly varying performance over time and across Justices. To support the case for a model’s future applicability, it should consistently outperform a baseline comparison.  

Both legal scholars and practicing lawyers have had difficulty leveraging prediction models \cite{bib5}. Among other difficulties, qualitatively-oriented legal experts tend to suggest model improvements based on anecdote and/or their own untested mental model. However, if these ostensible improvements cannot be systematically inferred from data, or if their impact on the model is detrimental in other periods or for other Justices, then they ought not be included in a model engineered for consistency.  

While prediction models can be applied in many contexts, consistency can also be related to a risk preference in a repeated betting scenario.  For example, instead of preferring the highest per-wager expected value (i.e., maximum accuracy), a bettor might prefer a wager with less volatility or long-term downside risk.  

Both consistency and generality can be seen as related to overfitting and the bias-variance trade-off.  But in addition to the typical learning problems under a stationary system, we are faced with a more complex reality.  Court outcomes are potentially influenced by a variety of dynamics, including public opinion as in \cite{bib9}, inter-branch conflict \cite{bib10}, both changing membership and shifting views of the Justices as explored in \cite{bib11} \cite{bib12}, and judicial norms and procedures \cite{bib13}. The classic adage “past performance does not necessarily predict future results" is very much applicable.  For example, likely due to changes in norms, the number of cases per term has fallen from approximately 150 between 1950-1990 to less than 90 between 1990-2015. Consider another famous historical example, as explored in \cite{bib14} and \cite{bib15}, when the aftermath of President Franklin D. Roosevelt’s attempted Court-packing plan in 1937 resulted in a significant turnover of Justices in years that followed.  Each of these and other changes represents a challenge to a model engineered with consistency as a goal.  

\subsection*{Out-of-Sample Applicability}
Our third model principle is out-of-sample applicability. Namely, all information required for the model to produce an estimate should be knowable prior to the date of decision. This is in contrast with models like \cite{bib2}, which require partial knowledge about the outcome to predict the full outcome. This principle is arguably the most important, as it allows for the model to generate predictions in advance, i.e., predictions that can be applied usefully in the real world.   

While existing approaches like \cite{bib1}, \cite{bib2} and \cite{bib3} may honor one or two of these principles, none simultaneously achieve all three above, severely limiting their general applicability. Both \cite{bib1} and \cite{bib3} are predictive out-of-sample but fail to be general enough to apply widely or consistent when tested. By contrast, \cite{bib2} is general across terms and consistent, but not predictive out-of-sample since it requires knowledge of some votes to predict others.  As detailed further below, our approach is the first that satisfies all three of these criteria, and thus represents a significant advance in the science of quantitative legal prediction. 

\section*{Data and Feature Engineering}

\subsection*{SCDB}
In order to build our model, we rely on data from the Supreme Court Database (SCDB)\cite{bib16}. SCDB features more than two hundred years of high-quality, expertly-coded data on the Court’s behavior.  Each case contains as many as two hundred and forty variables, including chronological variables, case background variables, justice-specific variables, and outcome variables.  Many of these variables are categorical, taking on hundreds of possible values; for example, the `\textsc{issue}' variable can take 384 distinct values.  These SCDB variables form the basis for both our features and outcome variables. 

SCDB is the product of years of dedication from Professor Harold Spaeth as well as many others. The database has been consistently subjected to reliability analysis and has been used in hundreds of academic studies (e.g., ~\cite{bib11}, ~\cite{bib17}, ~\cite{bib18}, ~\cite{bib19}, ~\cite{bib20}, ~\cite{bib21},  ~\cite{bib22}, ~\cite{bib23}). While there are serious and important limits to SCDB, as detailed in \cite{bib24}, SCDB is the highest-quality and longest-duration database for Supreme Court decisions.

There are currently two releases of SCDB: SCDB Modern and SCDB Legacy. The SCDB Modern release contains terms beginning in 1946, while the SCDB Legacy release contains terms beginning in 1791.  When \cite{bib25}, an earlier pre-print version of this paper was released, SCDB Legacy had not yet been released. As SCDB Legacy represents more than a threefold increase in the length of simulation history and size of training data, we have re-run all model construction and analysis for the new data release; methods and results from \cite{bib25} are thus superseded by this paper. 

\subsection*{Targets}
To model Supreme Court decisions, we need to define an outcome variable from SCDB corresponding to a decision. Typically, Court-watchers frame decisions as either affirming or reversing a lower court’s decision. This, however, is only consistent with cases heard on appeal. In some circumstances, the United States Supreme Court is the court of original jurisdiction, and there is therefore no lower court against which to frame reversal. In these cases, decisions are typically framed as either siding with the plaintiff(s) or defendant(s). In addition, the Court and its members may take technically-nuanced positions or the Court's decision might otherwise result in a complex outcome that does not map onto a binary outcome.
  
In order to build a general model that can handle all cases, we created a disposition coding map that defines a Justice vote as (i) Reversed, (ii) Affirmed, or (iii) Other, depending on a Justice’s vote and the SCDB's \textsc{caseDisposition}' variable.  This disposition coding map is outlined in our \textit{Github} repository \cite{bib26}. Our mapping displays Justice vote values by column and Court `\textsc{caseDisposition}' values by row.  The case outcome is defined as Reverse if there are more total Reverse votes than Affirm votes; notably, Other votes, which may include recusals or non-standard form decisions, are excluded from the vote aggregation. Table \ref{outcome_distribution} below displays the distribution of Reverse, Affirm, and Other coding by Justice outcome and case outcome.

\begin{table}[!ht]
\centering
\caption{{\bf Outcome Distribution (1816-2015)}}
\begin{tabular}{|r|c|c|c|}
\hline
\bf{Class} & \bf{Justice} & \bf{Case}\\\hline
Affirm & 113,454 & 16,718\\\hline
Reverse & 93,161 & 11,291\\\hline
Other & 37,267 & NA\\\hline \hline 
Total & 243,882 & 28,009\\\hline
\end{tabular}
\label{outcome_distribution}
\end{table}

\subsection*{Features and Feature Engineering}
With the outcome variable specified, we proceed next to describe the SCDB features used and feature engineering we performed. SCDB contains a wide range of potential features, and the majority of these are categorical variables. In our study, we begin with the following features available from SCDB: \textsc{Justice (ID)}, \textsc{term}, \textsc{natural court}, \textsc{month of argument}, \textsc{petitioner}, \textsc{respondent}, \textsc{manner in which Court took jurisdiction}, \textsc{administrative action}, \textsc{court of origin} and \textsc{source of the case}, \textsc{lower court disagreement}, \textsc{reason for granting cert}, \textsc{lower court disposition}, \textsc{lower court direction}, \textsc{issue}, and \textsc{issue area}. For each of these variables, we follow standard practice and convert the categorical variables into binary or indicator variables.  For example, in the case of reason for granting cert, there are 13 categories used in SCDB. Therefore, the single `\textsc{certReason}' variable is converted to 13 binary or indicator variables - one for each possible option. 

In addition to simple feature encoding, we also engineer features that do not occur in SCDB as released.  The first set of features that we engineer are related to the Circuit Court of Appeals from which the dispute arose. SCDB codes this data in the form of the case source and case origin, where the source corresponds to the opinion under review and the origin corresponds to the location of original filing. While there are over 130 unique courts that these variables may be coded as, scholars primarily group them by Circuit; Circuits have been shown to be a strong predictor of reversal during certain periods, as shown in \cite{bib27}.  Based on this guidance, we therefore developed a translation from each SCDB court ID to the corresponding Circuit.  The coding maps from these origin and source courts to a new set of 16 categorical values, which are then binarized as the raw features above.


The features engineered above can both be described as coarsened or collapsed.  We move on next to features that are derived through arithmetic or interaction of one or more features. The first of this class is a set of chronologically-oriented features related to oral argument and case timing. These features include (i) whether or not oral arguments were heard for the case, (ii) whether or not there was a rehearing, and (iii) the duration between when the cased was originally argued and a decision was rendered. These features are based on the qualitative observation that the length of time between argument and decision is related to the unanimity of the Court; for example, in the past three terms, the ten “fastest" decisions of each term have nearly all been unanimous 9-0. 

Item (iii) may seem at first to include future or out-of-sample knowledge.  However, in practice, the predictions for a case may evolve as new information about the case is acquired prior to the decision being rendered. For example, when the Court announces that a case will have arguments heard, the delay feature may be set to zero initially. Once the argument date passes, the delay feature is then incremented periodically. After each time step that passes, the feature matrix for undecided cases is updated, and the resulting predictions may therefore change.  Consistent with “online” learning approaches such as \cite{bib7} and \cite{bib8}, this does not require out-of-sample information; it only requires that the data and algorithm be re-run on a specified time interval for any undecided cases in a term. 
  
Lastly, we engineer features that summarize the “behavior" of a Justice, the Court, the lower court, and differences between them. These features fall into three categories: (i) features related to the rate of reversal, (ii) features related to the left-right direction of a decision, and (iii) features related to the rate of dissent. These features can be thought of as conditional empirical probabilities. For example, (i) includes, at a given term and for a given justice, the historically-observed proportion of votes to reverse.  Importantly, in addition to calculating these values for each justice, we also include difference terms between the Court as a whole and the individual justice.  These difference terms are, qualitatively, the relative inclination of a Justice to reverse compared to the Court. We repeat these calculations for other justice-specific features including direction and agreement features, providing quantitative measures of left-right political preference and rate of dissent. In addition, we include a difference term between the lower court’s decision direction and the Justice’s historically-observed mean direction; this provides a measure of how far apart, ideologically, the Justice is from the lower court's opinion on review (excepting original jurisdiction cases).  Together, these features provide relative information about Courts’ and Justices’ political and procedural leanings; for example, we find that reversal rates vary significantly even in the last 35 years at both the Court and Justice level.

\section*{Model Construction}
With features and outcome data defined, we proceed to discuss the construction of our model.  While this section provides a general overview of modeling procedures, readers interested in the technical details should review the \textit{Github} repository accompanying the paper,\cite{bib26}; all source code and data required to reproduce the results presented are freely available there. The model is developed in Python and all methods described below, unless otherwise indicated, are from scikit-learn 0.18\cite{bib28}. 

The modeling process begins by selecting a term $T^*$; in order to satisfy our three principles above, no information from term $T^*$ or after should be available during the training phase.  If we let each docket-vote feature vector $d_i$ and docket-vote outcome $v_i$ have term $T(d_i)$, then our training feature set for model term $T^*$ is $D_T = \{d_i : T(d_i) < T^*\}$ and our training target set $V_T$ corresponds to matching $v_i$ records.  While some information may be known \textit{intra-}term, i.e., for $\{d_i : T(d_i) = T^*\}$, this modeling procedure only retrains at the outset of each term. For example, while some decisions in term $T^*$ may have been observed by December, cases in January are predicted using only information prior to October. Other than the incremental delay feature discussed above, no information derived from the current court term is incorporated into the model until the start of the following term.

While we represent $D$ and $V$ above as sets of vectors, we can easily consider it to be a feature matrix with each docket-vote in a row and each feature in a column.  As of 2015, $D_{2015}$ based on SCDB Legacy (beta) has 249,793 docket-votes; under our feature engineering approach described above, $D_{2015}$ has 1,501 columns.  In many machine learning approaches, we might pre-process $D$ by rescaling, rotating, interacting, or removing columns.  Random forest classifiers, especially when applied to binarized or indicator variables, do not generally require pre-processing.  Furthermore, random subspace methods like random forests implicitly remove or ``select" features by subsetting the feature space for each sub-learner tree.  One weakness of the scikit-learn implementation of random forests relative to alternatives like xgboost, however, is its treatment of missing data.  In most cases, this is handled by mapping missing values to a separate ``missing'' indicator column during encoding; in some cases, however, a historical mean imputation may be used.  However, no additional feature selection or pre-processing methods are applied to $D$ prior to learning.

We next apply a learning algorithm to $D$ and $V$.  As noted previously, we selected a random forest classifier\cite{bib4}. Random forests are part of the family of ensemble methods. Ensemble methods leverage the wisdom of the statistical crowds. In the case of random forest classifiers, we construct a forest of statistically diverse trees using bootstrap aggregation on random substrates of our training data. To cast predictions, we simply calculate predictions for each of our individual trees and then average across the entire forest. While an individual statistical learner (a single tree) might offer an unrepresentative prediction of a given phenomenon, the crowdsourced average of a larger group of learners is often better able to forecast various sets of outcomes.  By generating many different decision trees with diverse information sets and then averaging over the results, ensemble methods can convert a set of otherwise weak learners into a collectively strong learner.

Not only have random forests proven to be ``unreasonably effective'' in a wide array of supervised learning contexts \cite{bib29}, but in our testing, random forests outperformed other common approaches including \textit{support vector machines} (LibLinear, LibSVM) and feedforward artificial neural network models such as 
\textit{multi-layer perceptron} implemented with \cite{bib30}.  For details of the implementation, interested readers are directed to the scikit-learn documentation \cite{bib28} and \cite{bib31} in particular.  

Of some note, however, is our experimentation with the $warm\_start$ parameter to ``grow'' the forest online.  Recall that at the beginning of each term, the model is retrained to incorporate newly observed data.  In \cite{bib25}, we built a ``fresh'' forest model each term with number of trees selected by cross-validated hyperparameter search.  In these published results, however, we have simulated performance using both ``fresh'' forests and ``growing'' forests, in which trees are added to an existing forest. Only under certain circumstances, such as the changing of the natural court; following the addition or loss of a Justice,  does the model build a ``fresh forest''.  For example, the models used to produce this paper's results were trained with 125 initial trees beginning in 1816 ($5 * 25$ trees, five for each term between 1791-1816).  Each term, in the absence of a natural court change, an additional five trees were trained and added to the prior term's forest.

Our implementation of this ``growing'' approach allows for substantially faster simulation times and more stable predictions, as it only need train a small number of trees per step.  Equally important is that most trees in the forest are stable for most years, and so the same inputs in year $T$ and $T+1$ are likely to produce the same predictions.

Generally speaking, most learners benefit from joint cross-validation and hyperparameter search.  For the ``fresh'' forest approach, in which a new random forest is built each term, we performed a number of experiments by grid-searching the number of trees, minimum number of leaves per node, maximum depth per tree, heuristic used to select the number of features per tree (e.g., $log, sqrt$), and split criterion (e.g., Gini vs. entropy) \textit{for each model retraining}, i.e., \textit{for each term}.  This approach allows the parameters to adapt over the nearly 200 years of change in historical sample composition and size.  However, we found that the marginal improvement in accuracy and F1 were not worth the substantial increase in computational requirement and decreased stability of predictions. In the simple examples included in the \textit{Github} repository, a cross-validated hyperparameter does not have a noticeable impact on accuracy over ``default'' random forest configuration. 

As a whole, our model construction applies standard pre-processing and learning approaches within each step, but experiments with purposeful and atypical design around longitudinal model application.  For simplicity of subsequent presentation and replication, only the ``growing'' forest approach described above with five trees per step is presented. All source and results are available at \cite{bib26} for a reader interested in the details of model specification and implementation. 

\section*{Model Testing and Results}

  The data and model described above allow us to simulate out-of-sample performance for nearly 200 terms at the Supreme Court. However, there is no single approach to assessing performance in this context.  Below, we present standard, un-adjusted machine learning diagnostic results derived from the application of our prediction model.  We present both results at the justice level (\textit{i.e.}, our performance on predicting the votes of individual justices) and our performance at the case level (\textit{i.e.}, predicting the overall outcome of the Court).  Then, we compare our accuracy to that of several potential ``null'' or ``baseline'' models. 

\subsection*{Performance of Case and Justice Prediction Model}
\subsubsection*{Justice Level Prediction Results}
To begin, we present the results of our Justice vote prediction model.  Recall that the Justice-level model predicts whether the vote will fall into three classes (Affirm, Reverse, Other), but that the outcome at the case-level depends on whether or not a given Justice's votes are Reverse or not.  As a result, in Table \ref{justice_performance_unadjusted}  and Table \ref{justice_reverse_performance_unadjusted} below, we present precision, recall, and F1 results for both three-class and two-class problems in the tables below. In total, over the period from 1816-2015, our model exhibits accuracy of 71.9\% at the Justice vote level. 

\begin{table}[!ht]
\centering
\caption{{\bf Justice-vote performance (three-class), un-adjusted assessment}}
\begin{tabular}{|r|c|c|c|c|c|}
\hline
\bf{Class} & \bf{Precision} & \bf{Recall} & \bf{F1-score} & \bf{Support}\\\hline 
Affirm & 0.61 & 0.79 & 0.69 & 113,666\\\hline
Reverse & 0.64 & 0.48 & 0.55 & 93,569\\\hline
Other & 0.84 & 0.59 & 0.69 & 39,540\\\hline \hline
Mean/Total & 0.66 & 0.64 & 0.64 & 246,775\\\hline
\end{tabular}
\label{justice_performance_unadjusted}
\end{table}

\begin{table}[!ht]
\centering
\caption{{\bf Justice-vote performance (two-class), un-adjusted assessment}}
\begin{tabular}{|r|c|c|c|c|c|}
\hline
\bf{Class} & \bf{Precision} & \bf{Recall} & \bf{F1-score} & \bf{Support}\\\hline
Not Reverse & 0.73 & 0.84 & 0.78 & 153,206\\\hline
Reverse & 0.64 & 0.48 & 0.55 & 93,569\\\hline\hline
Mean/Total & 0.69 & 0.70 & 0.69 & 246,775\\\hline
\end{tabular}
\label{justice_reverse_performance_unadjusted}
\end{table}

\subsubsection*{Case Level Prediction Results}
An alternative but related prediction task is the prediction of case outcomes.  While better understanding the behavior of Justices is of interest to some court observers, the prediction of case outcomes is the key capability that motivates litigants and can move markets \cite{bib32}. Table \ref{case_performance_unadjusted} presents case-level results from our prediction model.  The predicted case outcome is determined from whether or not the majority of individual Justice votes favor reversing the prior status quo. Starting in 1816 and carrying through the conclusion of the October 2014 term, our model correctly predicts 70.2\% of the Court's decisions. 

\begin{table}[!ht]
\centering
\caption{{\bf Case prediction performance, un-adjusted assessment}}
\begin{tabular}{|r|c|c|c|c|c|}
\hline
\bf{Class} & \bf{Precision} & \bf{Recall} & \bf{F1-score} & \bf{Support}\\\hline
Not Reverse & 0.71 & 0.83 & 0.77 & 16,748\\\hline
Reverse & 0.67 & 0.50 & 0.57 & 11,340\\\hline
Mean/Total & 0.70 & 0.70 & 0.69 & 28,080\\\hline
\end{tabular}
\label{case_performance_unadjusted}
\end{table}

\begin{figure}[!h]
\includegraphics[width=5.20in]{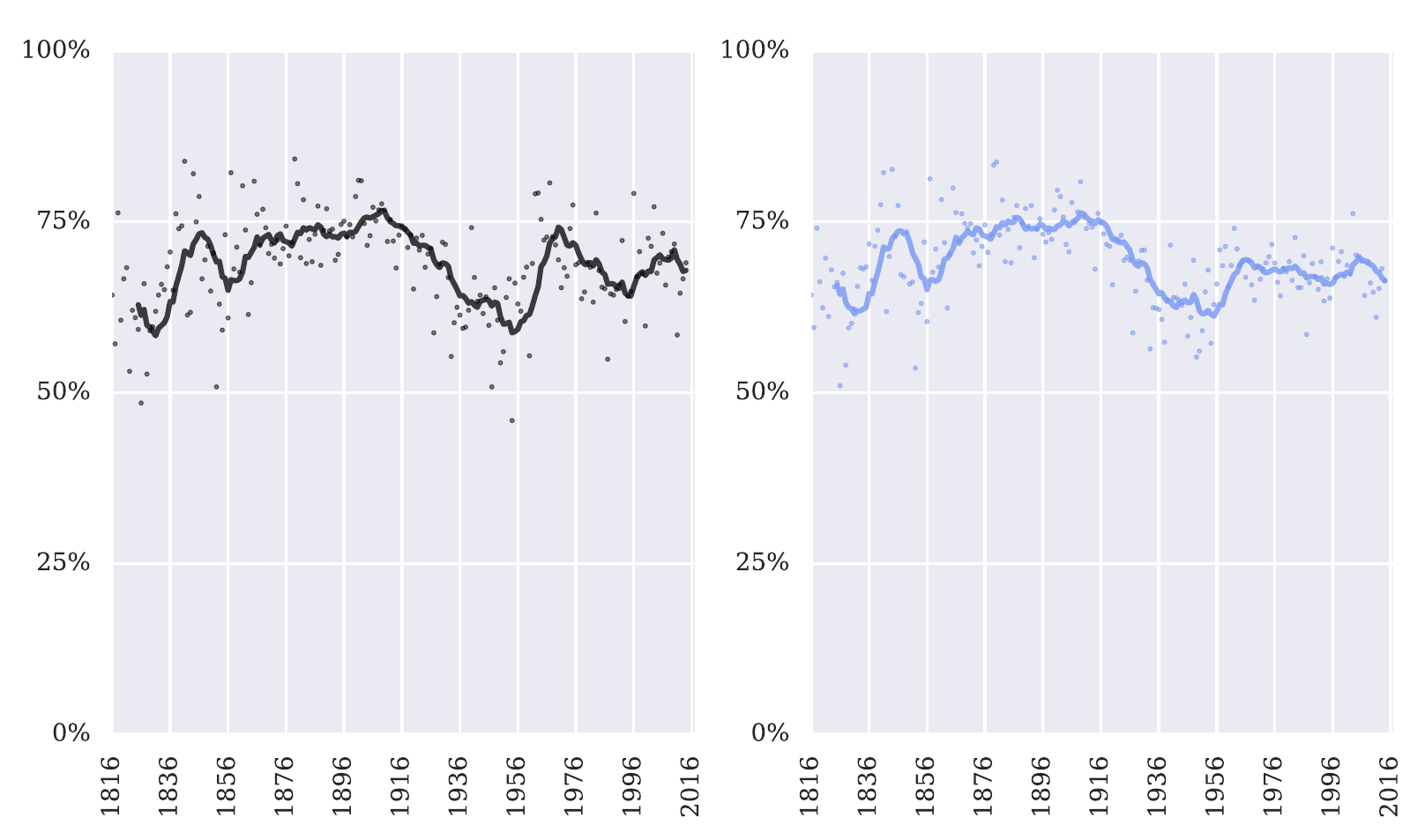}
\\
\caption{{\bf Case and Justice Accuracy 1816-2015 (by Term)} Time series of the accuracy of our prediction model at both the case level (left pane) and justice level (right pane)}
\label{figure_term_accuracy_comparison}
\end{figure}

Tables \ref{justice_performance_unadjusted}, \ref{justice_reverse_performance_unadjusted} and \ref{case_performance_unadjusted} provide the overall performance of our model (1816-2015). Figure \ref{figure_term_accuracy_comparison}, by contrast, demonstrates the consistency and generality of our approach over nearly two hundred years at both the case and justice level.  While some years and some decades are better than others, our model typically delivers stable performance for both cases outcomes and the votes of individual justices.

\subsection*{Candidate Baseline (Null) Models}
\begin{figure}[!h]
\includegraphics[width=4.65in]{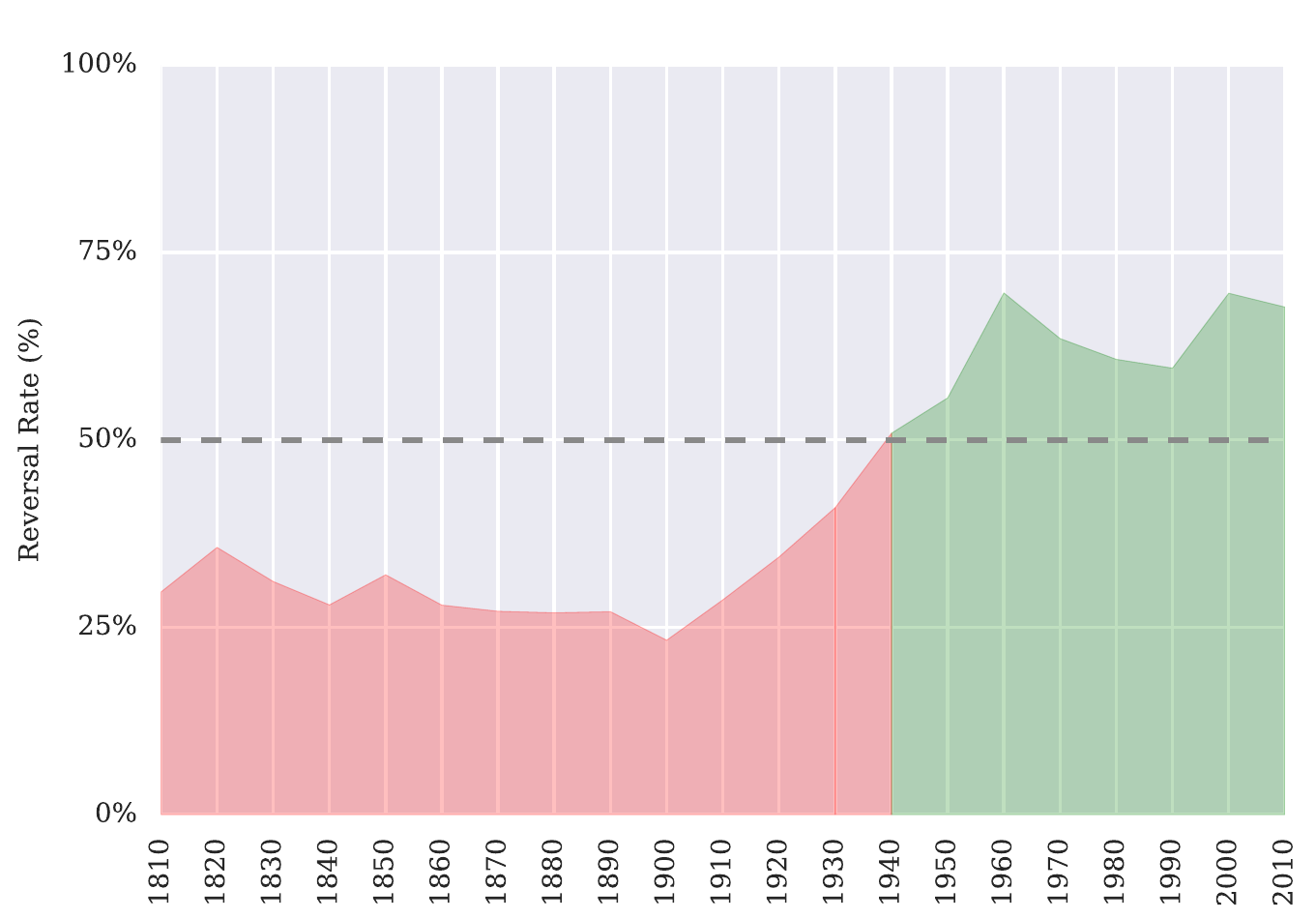}
\\
\caption{{\bf Reversal Rate by Decade}  For most of the Court's history, Reversal was much less frequent than it is now.  Only in recent history has Reversal become the more common outcome.}
\label{figure_reversal_rate_decade}
\end{figure}

But are the results above ``good?'' To meaningfully answer this question requires the development of a plausible baseline or null model. Specifically, while our approach may outperform an unweighted coin flip for both the two-class and three-class problems (50\% and 33\%, respectively), few legal experts would rely on an unweighted coin as a null model against which to compare their predictions.  Instead, informed by recent years, common wisdom among the legal community is that the baseline betting strategy should be to \textit{always guess Reverse}.  This strategy is supported by the recent history of the Court  over the last 35 terms: 57\% of Justice votes and 63\% of case outcomes have been Reverse.  However, this wisdom is quickly drawn into question when a broader view of history is taken into account, as Figure \ref{figure_reversal_rate_decade} demonstrates below. This trend is even more unbalanced when one considers the significant reduction in docket size over the past few decades, resulting in even more Affirm observations in previous years.

Since common wisdom appears too myopic to use as a historical baseline, we instead propose two additional null models to also use as comparisons. Specifically, in addition to the \textit{always guess Reverse} heuristic, consider two simple and similarly-intentioned rules: a most-frequent guessing strategies with an ``infinite'' memory and another with a ``finite'' memory.  The infinite memory baseline model, for a term $T$, simply guesses the most frequent outcome as observed in $D_T$. This model is most aligned with the spirit of common wisdom; however, as seen in Table \ref{outcome_distribution}, it results in a model that would still predict Affirm for the modern Court, and it has therefore significantly underperformed for the last 50 years.  In fact, at the current rate of dockets per year, it would take multiple decades worth of unanimous 9-0 decisions before this model would switch to predicting Reverse.

Therefore, we instead focus on an adapted most-frequent model featuring a ``finite window'' or ``moving average.'' Instead of determining the most frequent outcome over all history up to term $T$, only cases decided within the last $M < T$ terms are used. 

This memory parameter $M$ introduces a common hyper-parameter into the model definition. The optimization of ``memory'' parameters is a frequent challenge in many learning situations, especially ``online.''  In less technical terms, the optimization of $M$ can be reframed as a simple question: how much of the past is useful for predicting the future?  As is demonstrated by Figure \ref{figure_reversal_rate_decade}, it is often unclear when one should change strategy as underperformance is experienced.  It should be noted that this issue affects not just models in machine learning, but especially individual human experts attempting to leverage their personal experience and mental models.

While it is not possible to learn the optimal size of $M$ for all future states of the world, in our experiments, reproduced in our \textit{Github} repository\cite{bib26}, we have settled on a value of $M=10$. Not only does $M=10$ provide an easily-understood ``prior decade'' baseline, but also by selecting this as our memory window, we are able to test our prediction model against a null model built upon a value of $M$ that is nearly globally optimal for case accuracy. In other words, as $M=10$ is essentially derived by optimizing the $M$ hyperparameter \textit{in-sample}, i.e., using ``future information,'' this further advantages the baseline model and substantially hampers our efforts to outperform the null.  Despite this challenge, as demonstrated below, we still outperform the optimized baseline model over the past two centuries. 

Tables 5, 6 and 7 present the results from the justice and case-level for the $M=10$ ``finite'' memory null model.  Similar to the results reported for prediction model, Table \ref{justice_performance_baseline} displays the performance for the $M=10$ at the case level.  As above, the predicted case outcome is determined from whether the individual Justice votes are Reverse or Not Reverse.  In sum, optimizing the finite memory window using \textit{in sample} information yields justice-level accuracy of 66.2\% and case-level accuracy of 67.5\% from 1816-2015.  

\begin{table}[!ht]
\centering
\caption{{\bf Justice-vote performance (three-class), baseline model assessment}}
\begin{tabular}{|r|c|c|c|c|c|}
\hline
\bf{Class} & \bf{Precision} & \bf{Recall} & \bf{F1-score} & \bf{Support}\\\hline
Other & 0.33 & 0.01 & 0.02 & 39,540\\\hline
Affirm & 0.52 & 0.77 & 0.62 & 113,666\\\hline
Reverse & 0.57 & 0.46 & 0.51 & 93,569\\\hline\hline
Mean/Total & 0.50 & 0.53 & 0.48 & 246,775\\\hline
\end{tabular}
\label{justice_performance_baseline}
\end{table}

\begin{table}[!ht]
\centering
\caption{{\bf Justice-vote performance (two-class), baseline model assessment}}
\begin{tabular}{|r|c|c|c|c|c|}
\hline
\bf{Class} & \bf{Precision} & \bf{Recall} & \bf{F1-score} & \bf{Support}\\\hline
Not Reverse & 0.70 & 0.78 & 0.74 & 153,206\\\hline
Reverse & 0.57 & 0.46 & 0.51 & 93,569\\\hline\hline
Mean/Total & 0.65 & 0.66 & 0.65 & 246,775\\\hline
\end{tabular}
\label{justice_reverse_performance_baseline}
\end{table}

\begin{table}[!ht]
\centering
\caption{{\bf Case prediction performance, baseline model assessment}}
\begin{tabular}{|r|c|c|c|c|c|}
\hline
\bf{Class} & \bf{Precision} & \bf{Recall} & \bf{F1-score} & \bf{Support}\\\hline
Not Reverse & 0.69 & 0.81 & 0.75 & 16,740\\\hline
Reverse & 0.63 & 0.47 & 0.54 & 11,340\\\hline
Mean/Total & 0.67 & 0.67 & 0.66 & 28,080\\\hline
\end{tabular}
\label{case_performance_baseline}
\end{table}

\subsection*{Comparison against Baseline Models}

Above, we described three separate baseline models against which comparison might be undertaken: (1) the \textit{always guess Reverse} model, (2) the infinite memory model, $M=\infty$, and (3) the optimized finite memory model, $M=10$. At both the case and justice level, Figure 3 compares our prediction model to each of these null models. The left column corresponds to case accuracy, and the right column corresponds to justice accuracy.  The first row corresponds to $M=10$, the second row corresponds to $M=\infty$, and the third row corresponds to \textit{always guess Reverse}.  When our model outperforms the baseline, the plot is shaded green; when it fails to exceed the baseline performance, the plot is shaded red.

\begin{figure}[!h]
\includegraphics[width=5.7in]{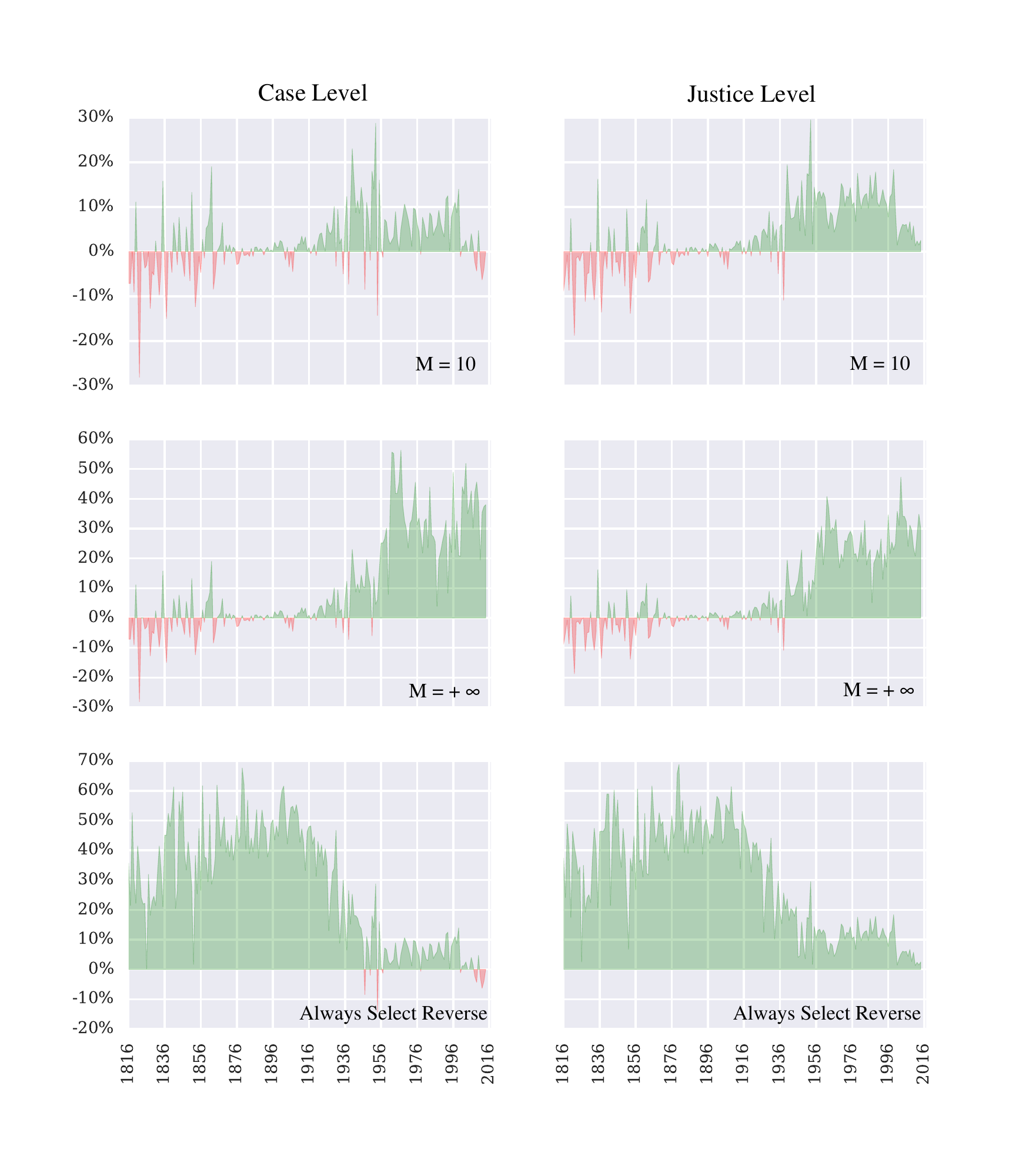}
\caption{ \textbf{Case and Justice Accuracy Compared Against Null Models} The first row corresponds to $M=10$, the second row corresponds to $M=\infty$, and the third row corresponds to \textit{always guess Reverse}. The left column corresponds to case accuracy, and the right column corresponds to justice accuracy.  When our model outperforms the baseline, the plot is shaded green; when it fails to exceed the baseline performance, the plot is shaded red.}
\label{figure_term_justice_accuracy}
\end{figure}

With respect to justice-level prediction, even a cursory review of Figure 3 demonstrates that our model performs very well against all baseline models across most of the last two centuries. Our model also performs well on the case-level predictions. Our approach especially outperforms both the \textit{always guess reverse} heuristic and the \textit{infinite} memory window during large, sustained periods.

After more than a century of soundly defeating all three null models, the performance of our prediction model has dipped during in the Roberts Court (as compared against the \textit{always guess reverse} heuristic and \textit{M=10} null model). Within the scope of this study, it is difficult to determine whether this represents some sort of systematic change in the Court’s macro-dynamics. However, thus far, it does appear that the Roberts Court is less predictable than its immediate predecessors.   

Flattening the data by taking each term as the relevant unit of analysis, Figure 4 offers an alternative perspective on our performance.  Figure 4 scores each term by comparing our performance to that of the null model. We assign a score of +1 to term where our model outperforms the null model, -1 in any term where our model performs worse than null model and 0 for any term where our model and the null offer identical performance. Given the results previously displayed in Figure 3, we only consider the M=10 null model for purpose of this analysis. Figure 4 plots the cumulative score of this tally as a function of time.

\begin{figure}[!h]
\includegraphics[width=5.1in]{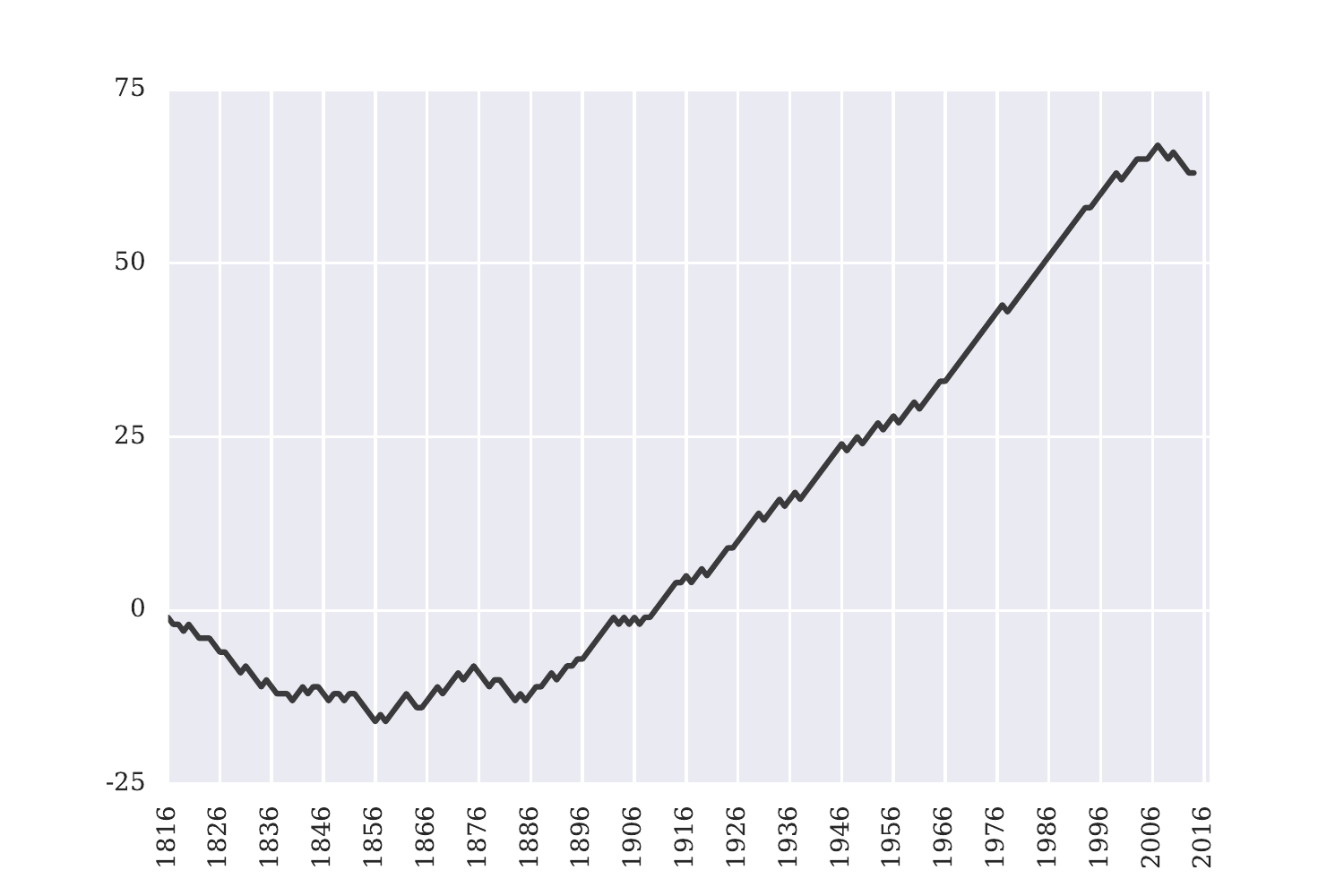}
\\
\caption{\bf {Cumulative Number of Terms Won Versus M=10 Null Model}}
\label{figure_term_justice_accuracy}
\end{figure}

A review of both Figure 3 and Figure 4 reveals that our prediction model initially struggles to outperform the $M=10$ finite memory null model. Several potential factors likely contribute to this. For example, as noted earlier, the $M=10$ null is \textit{in sample} optimized.  Thus, the derivation of the memory window through hyperparameter optimization is actually leveraging future information. By leveraging this class of future information, the \textit{in sample} optimization of \textit{M} appears to be better able than our model at fitting to some of the actual dynamics present in the early years of the Court.  

As the size of the training data increases, our model eventually surpasses the null. Namely, our twenty five year learning period (1791-1816) appears insufficient such that it requires several additional decades for our model to be able to consistently extract the signal from the noise.  In addition, the  ultimate success of our model vis-à-vis the null model is likely also driven by some increased level of behavioral stability on behalf of the Court starting in the second half of the nineteenth century. As reflected in Figure 4, starting after the conclusion of the American Civil War and in particular at the outset of the Fuller Court, our model begins to consistently outperform the \textit{in sample} optimized null model. 

\begin{figure}[!h]
\includegraphics[width=5.10in]{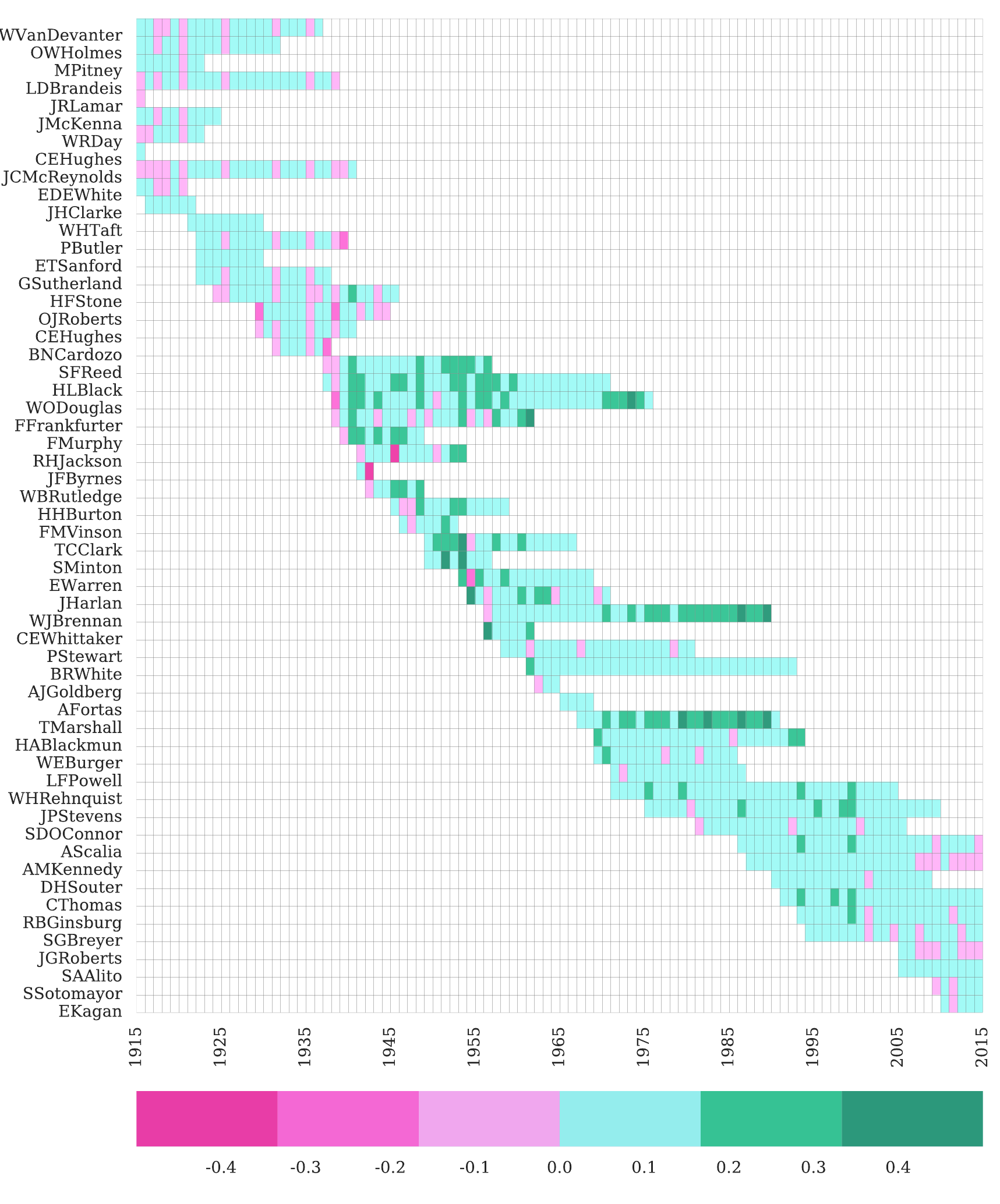}
\caption{ \textbf{Justice-Term Accuracy Heatmap Compared Against \textit{M=10} Null Model (1915 -2015)}. Green cells indicate that our model outperformed the baseline for a given Justice in a given term.  Pink cells indicate that our model only matched or underperformed the baseline.  The deeper the color green or pink, the better or worse, respectively, our model performed relative to the \textit{M=10} baseline.}
\label{Justice }
\end{figure}

Beyond performance on a term-by-term basis, another perspective on the performance of our model is to see how it performs on a justice-by-justice basis. At the justice-by-justice level over the past 100 years, Figure 5 displays our justice level performance against the most challenging of null models (i.e., the $M=10$ finite memory window). While careful inspection of Figure 5 allows the interested reader to explore the Justice-by-Justice performance of our model, a high level review of Figure 5 reveals the basic achievement of our modeling goals as described earlier.  While we perform better with some Justices than others and better in some time periods than others, Figure 5 displays generalized and consistent mean-field performance across many Justices and many historical contexts.   

\subsection*{Statistical Evaluation of Model Performance Against the Null Models}
While Figures 3, 4 and 5 as well as Tables 2 through 7 offer basic evidence regarding the performance of our model, we can now proceed to statistically measure the degree of confidence in our outperformance against the null. For completeness, in Table 8, we present the results of three tests for both our justice and case-level prediction models compared against the $M=10$ null model: (i) a paired $t$-test on annual case accuracy series, (ii) a Wilcoxon rank-sum test on annual case accuracy series, and (iii) a binomial test on per-case outcomes.  Tests (i) and (ii) evaluate whether, both under parametric and non-parametric assumptions, our model outperforms the baseline model at an aggregate, longitudinal level as measured by annual accuracy.  Test (iii), on the other hand, tests whether the distribution of individual model predictions is significantly better than a ``fairly''-weighted coin flip.  All tests are framed as one-sided tests that require our model accuracy to be greater than the null or baseline model. 

\begin{table}[!ht]
\centering
\caption{{\bf Summary of statistical tests: $p$-value}}
\begin{tabular}{|r|c|c|c|c|c|}
\hline
\bf{Test} & \bf{Justice (3-class)} & \bf{Justice (2-class)} & \bf{Case}\\\hline
Paired $t$-test & $O(10^{-58})$ & $O(10^{-11})$ & $O(10^{-5})$\\\hline
Wilcoxon rank-sum & $O(10^{-38})$ & $0.001$ & 0.03\\\hline
Binomial test & $\approx 0.0$ & $\approx 0.0$ & $O(10^{-19})$\\\hline
\end{tabular}
\label{statistic_summary}
\end{table}

These tests indicate that our random forest model significantly outperforms the baseline model, both at the aggregate, per-term level and at the per-case distribution. 



\section*{Conclusion and Future Research }

Building upon prior work in the field of judicial prediction \cite{bib1}, \cite{bib2}, \cite{bib3}, we offer the first generalized, consistent and out-of-sample applicable machine learning model for predicting decisions of the Supreme Court of the United States. Casting predictions over nearly two centuries, our model achieves 70.2\% accuracy at the case outcome level and 71.9\% at the justice vote level. More recently over the past century, we outperform an \textit{in-sample} optimized null model by nearly 5 \%. Among other things, we believe such improvements in modeling should be of interest to court observers, litigants, citizens and markets. Indeed, with respect to markets, given judicial decisions can impact publicly traded companies, as highlighted in \cite{bib32}, even modest gains in prediction can produce significant financial rewards. 

We believe that the modeling approach undertaken in this article can also serve as a strong baseline against which future science in the field of judicial prediction might be cast. While a researcher seeking to optimize performance for a given case or a given time period might pursue an alternative approach, our effort undertaken herein was to directed toward building a general model - one that could stand the test of time across many justices and many distinct social, political and economic periods.    

Beyond predicting U.S. Supreme Court decisions, our work contributes to a growing number of articles which either highlight or apply the tools of machine learning to some class of prediction problems in law or legal studies (e.g., ~\cite{bib5}, ~\cite{bib33}, ~\cite{bib34}, ~\cite{bib35}, ~\cite{bib36}, ~\cite{bib37},  ~\cite{bib38}, ~\cite{bib39}, ~\cite{bib40}).  We encourage additional applied machine learning research directed to these areas and new areas where the application of predictive analytics might be fruitful.  

At its core, our effort relies upon a statistical ensemble method used to transform a set of weak learners into a strong learner.  We believe a number of future advancements in field of legal informatics will likely rely on elements of that basic approach.  Namely, our focus on statistical crowd sourcing actually foreshadows future developments in the field. Future research will seek to find the optimal blend of experts, crowds \cite{bib41} and algorithms as some ensemble of these three streams of intelligence likely will produce the best performing model for a wide class of prediction problems\cite{bib42}.  


%

\section*{Acknowledgments}
We would like to thank our reviewers and all of those who provided comments on prior drafts of this paper.

\nolinenumbers

\end{document}